This manuscript has been accepted for publication in Medical Physics on June 15, 2025. Please use the following reference when citing the manuscript.

**Development and *in silico* imaging trial evaluation of a deep-learning-based transmission-less attenuation compensation method for DaT SPECT**

Zitong Yu[1], Md Ashequr Rahman[1], Zekun Li[1], Chunwei Ying[1], Hongyu An[2], Tammie L.S. Benzinger[2], Richard Laforest[2], Jingqin Luo[3], Scott A. Norris[2], and Abhinav K. Jha[1,2]

[1]Department of Biomedical Engineering, Washington University, St. Louis, MO, USA

[2]Mallinckrodt Institute of Radiology, Washington University, St. Louis, MO, USA

[3]Department of Surgery, Washington University, St. Louis, MO, USA

Corresponding author:

Abhinav K. Jha

1 Brookings Dr, St. Louis, MO, 63105, USA

Email: a.jha@wustl.edu

First author:

Zitong Yu

Ph.D. student in training

1 Brookings Dr, St. Louis, MO, 63105, USA

Email: yu.zitong@wustl.edu

Financial support: National Institute of Biomedical Imaging and Bioengineering grant R01-EB031051, R01-EB031962, and R01-NS124789.

## ABSTRACT

**Background:** Quantitative measures of dopamine transporter (DaT) uptake in caudate, putamen, and globus pallidus derived from DaT-single-photon emission computed tomography (SPECT) images are being investigated as biomarkers to diagnose, assess disease status, and track the progression of Parkinsonism. Reliable quantification from DaT-SPECT images requires performing attenuation compensation (AC), typically with a separate X-ray CT scan. Such CT-based AC (CTAC) has multiple challenges, a key one being the non-availability of X-ray CT component on many clinical SPECT systems. Even when a CT is available, the additional CT scan leads to increased radiation dose, costs, and complexity, potential quantification errors due to SPECT-CT misalignment, and higher training and regulatory requirements.

**Purpose:** To overcome the challenges with the requirement of a CT scan for AC in DaT SPECT, we develop a transmission-less AC method for DaT SPECT and validate the method in a clinically realistic setting using an *in silico* imaging trial.

**Method:** Integrating concepts from physics and deep learning, we propose a deep learning-based transmission-less AC method for DaT-SPECT (DaT-CTLESS). In this method, an initial attenuation map reconstructed from scatter-energy window projection is segmented into different regions using a U-net-based network trained on CT scans. Each region is assigned a predefined attenuation coefficient, yielding an attenuation map for AC. An *in silico* imaging trial, titled ISIT-DaT, was designed to evaluate the performance of DaT-CTLESS on the regional uptake quantification task. In this trial, DaT SPECT scans of a virtual patient population, curated from CT and MR images of real patients, were generated with simulated SPECT scanners from two vendors. The Society of Nuclear Medicine guidelines suggest using uniform attenuation map for AC (UAC) when a CT scan is unavailable. Thus, the primary objective of ISIT-DaT was to assess whether the correlation between the activity uptake obtained using DaT-CTLESS and CTAC, as quantified using the intraclass correlation coefficient (ICC), was higher than the correlation between UAC and CTAC. Secondary objectives included evaluating DaT-CTLESS on the task of distinguishing patients with normal versus reduced DaT-specific binding ratio (SBR) of putamen, evaluating the repeatability of DaT-CTLESS in a test-retest study, assessing the generalizability across two SPECT scanners, evaluating DaT-CTLESS using fidelity-based figures of merit (FoMs), and evaluating the sensitivity of DaT-CTLESS to intra-regional uptake heterogeneity. Finally, we compared DaT-CTLESS with two other deep-learning transmission-less AC methods on regional uptake quantification across different training dataset sizes.

**Results:** In the ISIT-DaT trial, data from 150 virtual patients were used to train, and another 47 were used to evaluate the DaT-CLTESS method. We observed that DaT-CTLESS yielded a significantly higher correlation with CTAC than the correlation between UAC and CTAC on the regional DaT uptake quantification task. Further, DaT-CLESS had an excellent agreement with CTAC (ICC: 0.96, 95% CI: [0.94,0.97], p<0.05) on this task, significantly outperformed UAC in distinguishing patients with normal versus reduced putamen SBR and on fidelity-based FoMs, yielded good generalizability across two scanners, was generally insensitive to intra-regional uptake heterogeneity, demonstrated good repeatability in the test-retest study, exhibited robust performance even as the size of the training data was reduced, and generally outperformed the other considered deep learning methods on the task of quantifying regional uptake across different training dataset sizes.

**Conclusion:** The proposed DaT-CTLESS method, as evaluated in ISIT-DaT trial, was observed to yield reliable performance for transmission-less AC in DaT-SPECT, providing a strong motivation for further clinical evaluation.

**Keywords:** In silico imaging trial, dopamine transporter SPECT, transmission-less attenuation compensation, deep learning, task-based evaluation.

## INTRODUCTION

Parkinson disease (PD), the second-most common neurodegenerative disease, is expected to affect 12 million people worldwide by 2040[1]. To improve early and accurate diagnoses, assess disease status, track progression, and monitor therapeutic response, objective biomarkers are needed. In patients with neurodegenerative parkinsonism, abnormal changes in the dopamine transporter (DaT) concentration in the striatal region, including caudate, putamen, and globus pallidus (GP), are being investigated as biomarkers[2]. [123]I-Ioflupane has a high affinity to DaT, and this affinity depends on the availability of DaT in the striatal region, which relates to PD severity. DaT single-photon emission computed tomography (SPECT), a widely used clinical imaging study, provides a mechanism to quantify DaT uptake in the striatal region[3,4]. However, this quantification task is challenging due to the multiple image-degrading effects in SPECT, including the attenuation and scatter of photons, limited system resolution, and noise. The challenge of attenuation, which is the focus of this manuscript, is particularly severe for the caudate, putamen, and GP regions since they are located deep in the brain[5,6].

To address the image-degrading effects due to attenuation, multiple attenuation compensation (AC) methods for SPECT have been developed[7-9]. These AC methods require an attenuation map, typically obtained using a transmission scan, such as X-ray CT scan, and typically acquired on dual-modality SPECT/CT scanners. However, multiple SPECT systems do not have a CT component[10]. This includes SPECT systems in many community hospitals and most physician offices, mobile SPECT systems that enable imaging in remote locations, and many of the emerging solid-state-detector-based SPECT systems that provide higher sensitivity, as well as higher energy, temporal, and spatial resolution[11,12]. In fact, close to 90% of the scans in the Parkinson's Progression Markers Initiative (PPMI), one of the largest open-access SPECT databases for patients with PD, do not have a transmission scan, which limits the potential of using this database for quantitative analysis. Also, even when SPECT/CT systems are available, an additional CT scan leads to additional radiation dose, higher costs, and potential quantification errors caused by misalignment between the SPECT and the CT scans[13-15]. This misalignment may occur more frequently in patients with PD due to extraneous movement during scan[16,17]. For these reasons, there is an important need for transmission-less AC methods.

Given the high significance, several transmission-less AC methods have been proposed[18-22]. Early set of methods were based on modeling the physics of the SPECT emission data and can be classified into two categories. The first category of the methods involves the direct estimation of attenuation coefficients from SPECT emission data, by applying operation on iterative inversion of the forward model of the SPECT system[18,19] or by leveraging the consistency conditions grounded in the forward model[20]. In the second category, attenuation maps are estimated using emission data acquired in scatter windows. The methods within this category encompass approaches that assign attenuation coefficients to segments of an initial attenuation map estimate[21] and approaches based on inversion of scattering models[22,23]. However, these physics-based methods have exhibited limited accuracy[7].

More recently, deep learning (DL) has shown promise in transmission-less AC for SPECT[24-31]. While mostly such methods have been proposed for cardiac SPECT[24-28], DL-based transmission-less methods for brain SPECT have also been proposed[29-31]. These DL-based AC methods can be grouped into direct and indirect AC methods[25]. The indirect AC methods attempt to learn the attenuation map from non-AC images[24,31], while the direct AC methods learn the AC images directly from the non-AC images[29,30]. While these methods have shown promise, they assume a correlation between the non-AC images and attenuation distributions for indirect methods, or a correlation between non-AC and AC images for direct methods. However, the physical premise for such correlations remains unclear and may be violated. For example, when conducting brain SPECT, the head gear worn by patients leads to attenuation, but there is no activity uptake at the site of this attenuation.

In SPECT imaging, the dominant photon-interaction mechanism is Compton scattering. The Compton-scattered photons are recorded primarily in the scatter-energy window projections. These scattered locations also include the site of no activity uptake, such as the head gear. In a recent study, it was quantitatively observed that scatter-window projection data contains information to estimate the attenuation distribution[32]. In fact, AC methods have been proposed based on the premise that the probability of scatter at a location is directly proportional to the attenuation coefficient at that location[33]. One class of methods rely on estimating the attenuation coefficient for each voxel. Both physics[7,22,23] and DL-based[24,26] approaches have been proposed in this regard. However, the task of estimating each of the voxel values from the limited count scatter-window data is a high-dimensional ill-posed problem. A second class of methods include those that segment an initial scatter-window reconstruction into regions and assign predefined attenuation coefficients to estimate the attenuation map[21]. This reduces the dimensionality of the problem. However, the accuracy of segmentations yielded by these methods is limited. In this context, DL has been showing strong promise in image segmentation applications, including in SPECT[34-36]. Further, with SPECT, samples of SPECT emission data and corresponding CT scans are available, which may provide a mechanism for a DL-based approach to segment the initial scatter-window reconstruction. Based on this, we previously developed a transmission-less AC method for myocardial perfusion SPECT[37] and observed that the method yielded reliable performance on the task of perfusion defect detection[38]. Motivated by the positive findings in that study, we now extend the idea of integrating scatter-window data and deep-learning to propose a new transmission-less AC method for DaT SPECT (DaT-CTLESS).

As recommended in the best practices for evaluation of artificial intelligence (AI) algorithms for nuclear medicine (the RELAINCE guidelines)[39], we evaluate the DaT-CTLESS method on the clinical task of quantifying DaT uptake within the caudate, putamen, and GP regions. Evaluation of DL-based methods on clinically relevant estimation tasks requires the ground-truth activity values within the volumes of interest (VOIs). However, in clinical trials, ground truth is generally unavailable. While this absence of ground truth can be addressed by conducting physical phantom studies, these studies are constrained by their inability to comprehensively represent the patient populations. In addition, for clinical translation, it is necessary to assess generalizability and repeatability of DL-based imaging methods. In a clinical trial, this would usually involve administering radiotracers to the same patient multiple times and imaging them across different scanners. This process would pose substantial radiation exposure to patients, be expensive and time-consuming, and have multiple logistical challenges. In this context, the emerging virtual imaging trial (VIT) paradigm provides a mechanism to rigorously and objectively evaluate imaging technology in simulated clinical scenarios that address these challenges by providing the ability to simulate patient-population variability and imaging-system physics. Further, the same virtual patient can be scanned multiple times on different scanners in a VIT, overcoming the practical challenges in clinical trials. Moreover, because the ground truth is known, VIT enables to assess performance on the clinically relevant quantification tasks[40-42]. For example, Badano *et al.*, through the VICTRE *in silico* trial, envisioned that VITs have the potential of providing regulatory evidence for imaging technologies[41].

In this study, we designed an *in silico* imaging trial for DaT SPECT (ISIT-DaT) to evaluate the similarity between DaT-CTLESS and CTAC methods on regional DaT uptake quantification task in clinically realistic settings. We compared the DaT-CTLESS method with the AC method when CT is available (CTAC) and an AC method that uses a uniform attenuation map (UAC). The Society of Nuclear Medicine guidelines suggest that UAC can be used when a CT scan is unavailable[43]. Generally, the UAC method is clinically employed when CT scans are unavailable[44]. We hypothesized that on the task of regional DaT uptake quantification, DaT-CTLESS would achieve a higher correlation with CTAC compared to that between UAC and CTAC. In addition, we evaluated the performance of DaT-CTLESS on distinguishing patients with normal versus reduced putamen DaT specific binding ratio (SBR). Moreover, the generalizability of DaT-CTLESS was evaluated across two SPECT scanners and the repeatability was evaluated in a test-retest study. We also evaluated DaT-CTLESS using fidelity-based figures of merit (FoMs). Additionally, we evaluated the sensitivity of DaT-CTLESS to different levels of intra-regional uptake heterogeneity. Finally, we compared the DaT-CTLESS method to a direct and another indirect DL-based transmission-less AC method with different sizes of training dataset.

**METHODS**

**DaT-CTLESS method**

Consider a SPECT system imaging a tracer distribution $f(\mathbf{r})$ within a patient, where $\mathbf{r} \in \mathbb{R}^3$ denotes the 3-dimensional coordinates. The SPECT system yields the projection data in photopeak and scatter-energy windows, denoted by M-dimensional vectors $\boldsymbol{g}_{pp}$ and $\boldsymbol{g}_{sc}$, respectively. Denoting the reconstructed image by an N-dimensional vector $\hat{\boldsymbol{f}}$, the attenuation map by an N-dimensional vector $\boldsymbol{\mu}$, and the reconstruction operator with AC by $\mathcal{R}_{\boldsymbol{\mu}}$, we have

$$\hat{\boldsymbol{f}} = \mathcal{R}_{\boldsymbol{\mu}}(\boldsymbol{g}_{pp}). \tag{1}$$

We investigated the use of the SPECT projection data $\boldsymbol{g}$ to estimate the attenuation map $\boldsymbol{\mu}$. Denote the reconstruction operator without AC by $\mathcal{R}$ and the scatter-energy window reconstruction by $\hat{\boldsymbol{f}}_{sc}$, we have

$$\hat{\boldsymbol{f}}_{sc} = \mathcal{R}(\boldsymbol{g}_{sc}). \tag{2}$$

Similarly, the photopeak-energy window reconstruction without AC, denoted by $\hat{\boldsymbol{f}}_{pp}$, is given by

$$\hat{\boldsymbol{f}}_{pp} = \mathcal{R}(\boldsymbol{g}_{pp}). \tag{3}$$

Consider a segmentation operator denoted by $\mathcal{D}_{\boldsymbol{\Theta}}$, which is parameterized by $\boldsymbol{\Theta}$. This segmentation operator is trained to estimate attenuation-region segments, given $\hat{\boldsymbol{f}}_{pp}$ and $\hat{\boldsymbol{f}}_{sc}$ as input, i.e.,

$$\{\widehat{\boldsymbol{\Phi}}\} = \mathcal{D}_{\boldsymbol{\Theta}}(\hat{\boldsymbol{f}}_{pp}, \hat{\boldsymbol{f}}_{sc}), \tag{4}$$

where $\{\widehat{\boldsymbol{\Phi}}\} = \{\widehat{\boldsymbol{\Phi}}^1, \widehat{\boldsymbol{\Phi}}^2, \ldots, \widehat{\boldsymbol{\Phi}}^K\}$ is the set of estimated region segments.

We used a U-net with multi-channel input to perform this segmentation operator $\mathcal{D}_{\boldsymbol{\Theta}}$. The input of the encoder consists of two channels: one receives the scatter-window reconstructions, and the other receives the photopeak-window reconstructions. The encoder consists of seven convolutional layers with $3 \times 3 \times 3$ kernels and step sizes of 1 or 2, decreasing the size of images from $128 \times 128 \times 128$ to $16 \times 16 \times 16$. The decoder contains three transposed convolutional layers with $3 \times 3 \times 3$ kernels and step sizes of 2. After each convolutional layer in the encoder and each transposed convolution layer in the decoder, a leaky rectified linear unit is applied. Skip connections with attention gate are applied between the encoder and the decoder to improve segmentation performance[45]. In the final layer, a SoftMax function is applied, yielding the final estimated segmentation. The loss function of the segmentation network is the weighted cross-entropy loss between the estimated segmentations and the CT-derived segmentations.

The segmentation network was trained to estimate attenuation-region segments $\{\widehat{\boldsymbol{\Phi}}\}$, which included white/gray matters, skull, scalp, and the head gear. We assumed that the attenuation coefficient in each region was constant, denoted by a scalar value $\mu_k$, where $k$ is the region index. The attenuation coefficient $\mu_k$ is then assigned to corresponding estimated region segments. Denote the final estimate of attenuation map by $\widehat{\boldsymbol{\mu}}$, so that

$$\widehat{\boldsymbol{\mu}} = \sum_{k=1}^{K} \mu_k \widehat{\boldsymbol{\Phi}}^k . \tag{5}$$

This final estimate of the attenuation map is subsequently used for AC as in Eq. 1. The workflow of this approach, which we refer to as DaT-CTLESS, was shown in Fig. 1.

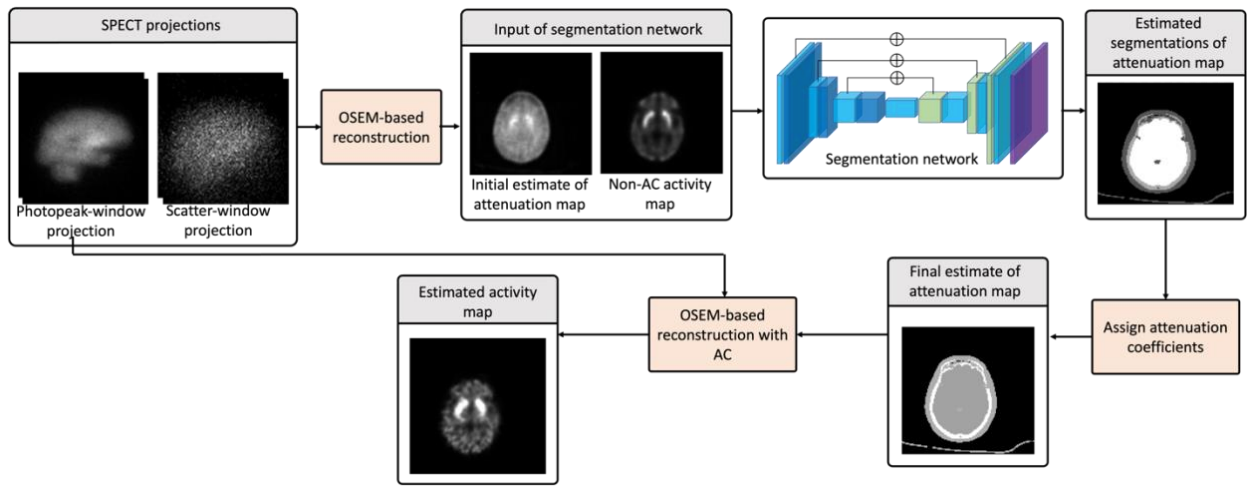

Figure 1. The workflow of quantifying regional DaT uptake using the DaT-CTLESS method.

**In Silico Imaging Trial Evaluation**

*Study Design*

We conducted the ISIT-DaT trial at Washington University in St. Louis to assess the performance of DaT-CTLESS on DaT uptake quantification within caudate, putamen, and GP regions. Fig. 2 shows a schematic of the trial design.

To conduct this trial, we collected images from participants (N = 197) enrolled in a neuroimaging study of memory and aging with an Institutional Review Board (IRB)-approved protocol and written informed consent. These participants included two sub-groups. The first sub-group of participants (group A, $N$ = 174) underwent single-time-point tri-modality imaging (PET/MR/CT), and the second sub-group of participants (group B, $N$ = 23) underwent dual-time-point tri-modality imaging. Thus, there were a total of 220 (174+23×2) tri-modality imaging studies in this trial. MR $T_1$-weighted images were acquired using an integrated Biograph mMR PET/MRI system (Siemens AG, Erlangen, Germany) and a 3D magnetization-prepared rapid gradient-echo (MP-RAGE) sequence. The parameters for the MR $T_1$-weighted images were as follows: TE/TR = 2.95/2300 ms, TI = 900 ms, image matrix = 240×256×170, and voxel resolution = 1.05×1.05×1.2 $mm^3$. CT images of these participants were also available, acquired using a Biograph 40 PET/CT system (Siemens AG, Erlangen, Germany) at 120 kVp with image size of 512×512×74 or 512×512×111 and voxel size of 0.59×0.59×3.0 $mm^3$ or 0.59×0.59×2.0 $mm^3$. Based on these MR and CT images, we generated 197 virtual patients with realistic ratios of striatal to occipital [123]I-ioflupane binding. These virtual patients were scanned by two realistically simulated clinical SPECT scanners, namely a GE Discovery 670 SPECT scanner and a Siemens Symbia Evo Excel SPECT scanner. We applied the DaT-CTLESS method, the CTAC method, the UAC method, and two other DL-based transmission-less AC methods to estimate the regional DaT uptake from those virtual patients.

The primary objective of ISIT-DaT was to assess whether the correlation between the activity uptake obtained using DaT-CTLESS and CTAC, as quantified using the intraclass correlation coefficient (ICC), was higher than that between UAC and CTAC. Secondary objectives included evaluating the performance of the DaT-CTLESS method on the task of distinguishing between patients with normal versus reduced putamen SBR values, evaluating the repeatability of DaT-CTLESS in a test-retest study, assessing the generalizability of DaT-CTLESS across two SPECT scanners, assessing DaT-CTLESS using fidelity-based FoMs, evaluating the sensitivity of DaT-CTLESS to intra-regional uptake heterogeneity, and comparing the DaT-CTLESS method with two other DL-based methods for different sizes of the training dataset.

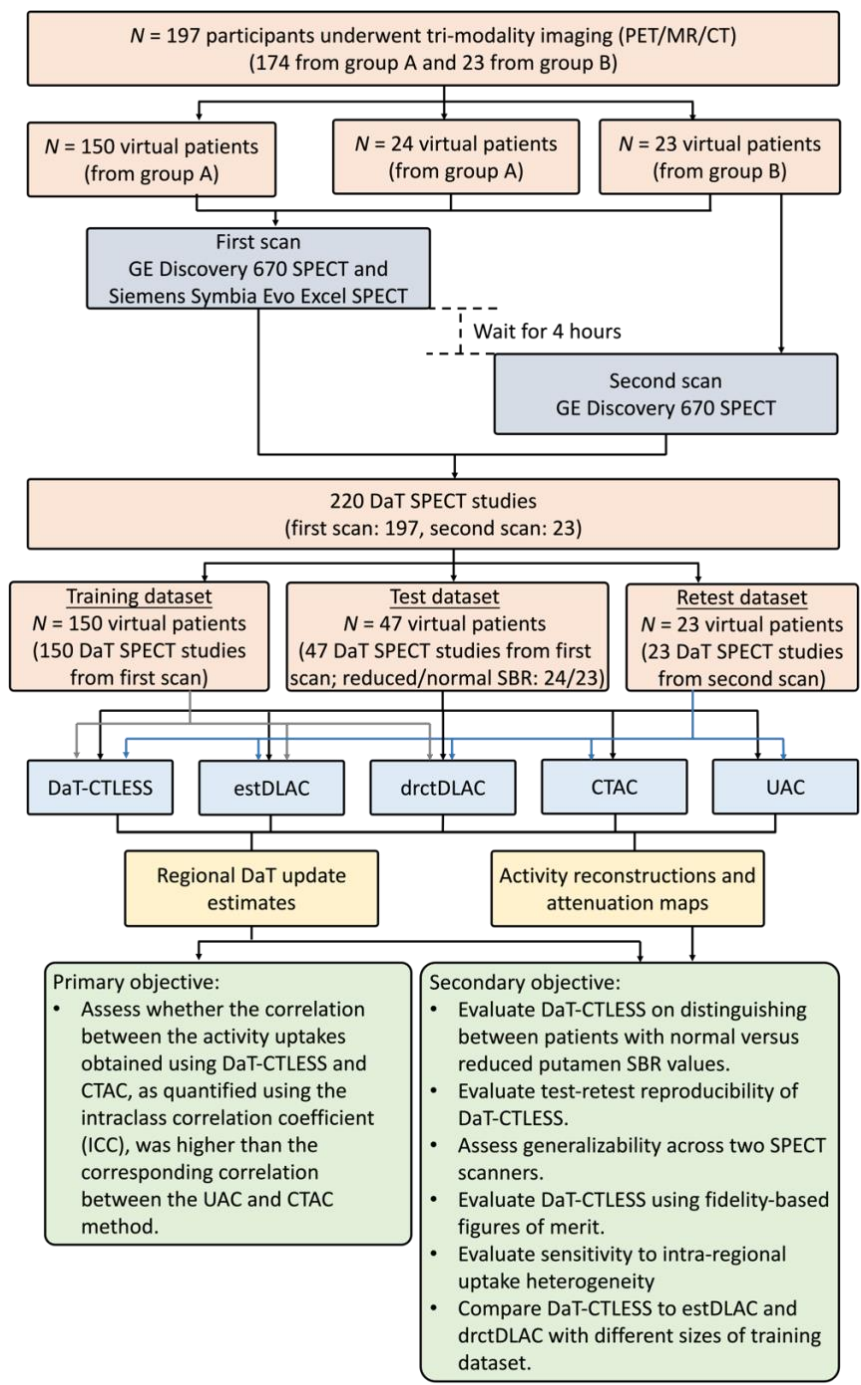

Figure 2. The overall design of the ISIT-DaT

*Trial Population*

The procedure of generating the trial population is shown in Fig. 3. We first registered the MR images with the corresponding CT images from collected tri-modality imaging studies using FMRIB's Linear Image Registration Tool (FLIRT)[46], a fully automated registration tool. As suggested in the Society of Nuclear Medicine (SNM) Practice Guideline for DaT SPECT[43], we considered VOIs in the striatum, including left and right caudates and putamen. We also considered left and right GPs as two additional VOIs. We segmented MR images into these six VOIs and other soft tissues using FreeSurfer software[47], yielding clinical-image-derived anatomical templates of the brain. To realistically simulate DaT uptake within these VOIs for each patient, SBR values within each VOI were independently sampled from Gaussian distributions. The mean and standard deviation values for the SBR distributions were derived from clinical data of $N$ = 45 PD patients and $N$ = 45 normal individuals[48]. By sampling the SBR values from these two clinically derived distributions, we simulated $N$ = 99 and N = 98 virtual patients with SBR values corresponding to those with PD and normal individuals, respectively. For conciseness, we refer to these patients as those with reduced SBR values and those with normal SBR values, respectively.

To generate images for training, the activity distribution within the various VOIs was considered homogeneous. However, while evaluating the method, we recognize that in patients with PD, intra-regional heterogeneity may be present. To study the performance of the method in the presence of intra-regional uptake heterogeneity, for patients with reduced SBR values, we modeled varying levels of heterogeneity in activity distributions within VOIs, using a similar approach as in a previous study[49]. More specifically, we modeled the heterogeneous distribution of activity uptake within caudate, putamen, and GP regions using a 3D lumpy model. Denote the support of region $k$ by $\phi_k(\mathbf{r})$, where $\mathbf{r} \in \mathbb{R}^3$ denotes the 3-dimensional coordinates. The heterogeneous uptake distribution of region $k$, denoted by $f_k(\mathbf{r})$, is given by

$$f_k(\mathbf{r}) = a_k \phi_k(\mathbf{r}) \sum_{p=1}^{P_k} \frac{1}{\left(\sigma_k \sqrt{2\pi}\right)^3} \exp\left[-\frac{\left(\mathbf{r} - \mathbf{c}_p\right)^2}{2\sigma_k^2}\right], \tag{6}$$

where $a_k$, $P_k$, and $\sigma_k$ denote the magnitude, total number, and width of lumps in the putamen region, respectively, and $\mathbf{c}_p$ denotes the center of the $p^{th}$ lump. Each $\mathbf{c}_p$ was randomly sampled from a uniform distribution within region $k$. The value of $a_k$ was chosen so that the SBR in region $k$ matched with that sampled from the clinical data of N = 45 PD patients[48]. The total number of lumps $P_k$ was sampled from a Poisson distribution with a mean of $\bar{P}_k$. Note that increasing the value of $\bar{P}_k$ and decreasing values of $\sigma_k$ would result in higher heterogeneity.

To simulate patients with varying levels of uptake heterogeneity, we used a mixture model. We first simulated fully heterogeneous patients with high uptake variation within the VOIs of putamen, caudate, and GP by considering a high number of lumps (setting $\bar{P}_k$ to 400) and a small lump width (setting $\sigma_k$ to 1.75 mm). We refer to these generated patients as fully heterogeneous patients. Denote the activity distribution in one of these patients by $f^{het}(\mathbf{r})$. We then simulate the same patient with the same SBR value but homogeneous uptake in the VOIs. Denote the activity distribution in one of these homogeneous patients by $f^{hom}(\mathbf{r})$. The different levels of heterogeneity were simulated using the following mixture model:

$$f_{\mathcal{W}}^{mix}(\mathbf{r}) = \mathcal{W}f^{het}(\mathbf{r}) + (1 - \mathcal{W})f^{hom}(\mathbf{r}), \tag{7}$$

where $\mathcal{W} \in \{0, 0.2, 0.4, 0.6, 0.8, 1\}$. A higher value of $\mathcal{W}$ led to a higher heterogeneity of uptake distribution. Fig. 3 shows activity maps of a representative patient with homogeneous activity uptake ($\mathcal{W} = 0$), mild heterogeneity ($\mathcal{W} = 0.2$), and fully heterogeneous activity uptake ($\mathcal{W} = 1$).

The attenuation maps were generated from CT scans by converting the Hounsfield units (HU) in CT images to linear attenuation coefficients in cm$^{-1}$ using a bi-linear model[50]. To generate attenuation-region masks for the training of DaT-CTLESS, a K-means clustering approach was used to segment the attenuation maps into attenuation regions, including white/gray matters, skull, scalp, and the head gear, based on attenuation coefficients[51].

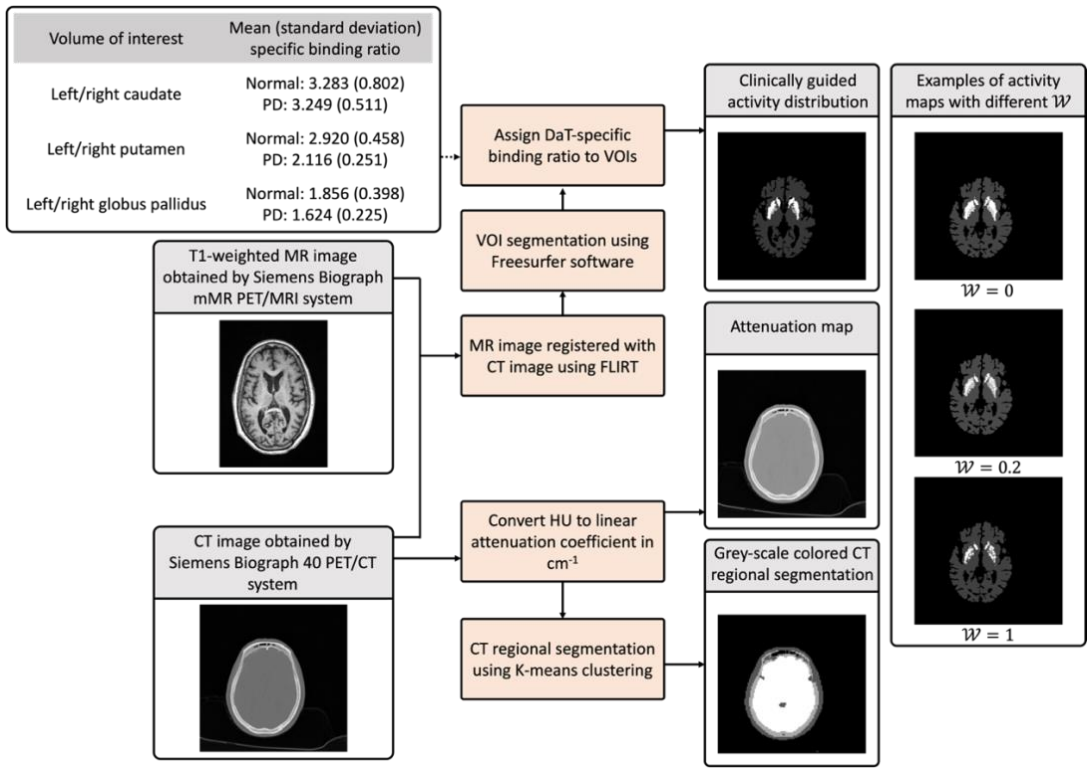

Figure 3. Virtual patient generation procedure

*Imaging Protocol*

We simulated a clinical scenario in which patients received a single intravenous injection of 5 mCi [123]I-Ioflupane, which is a typical dosage recommended in the SNM Practice Guideline. We assumed the ratio of striatal to occipital [123]I-ioflupane binding would be stable at 3 hours after injection of the radiotracer[52]. Thus, we chose this time point for performing DaT SPECT scanning.

SPECT scans were generated using SIMIND, a validated Monte Carlo-based SPECT simulation software[53]. We simulated two clinical SPECT systems with parameters similar to a GE Discovery 670 scanner with a Sodium iodide-based camera (GE Healthcare, Haifa, Israel) (GE-scanner) and a Siemens Symbia Evo Excel scanner (Siemens AG, Erlangen, Germany) (Siemens-scanner). The acquisition parameters of these two SPECT systems are listed in Table 1. The patients were imaged in a supine position with their head on an off-the-table head gear. A $128 \times 128$ acquisition matrix with a pixel size of 0.35 cm was used. As recommended in the SNM Practice Guideline, the scanning time was adjusted to ensure that 1.85 million total counts were detected within the photopeak window.

Following the first round of SPECT scan, $N$ = 23 patients generated from group B underwent a second round of scan four hours after the end of the first scan, simulating a test-retest study. The VOI masks and attenuation maps of these $N$ = 23 patients for the second scan were simulated based on the MR and CT images from tri-modality imaging studies taken at the second time points. Therefore, the VOI masks and attenuation maps were different from those in the first round of DaT SPECT scan, despite being derived from the same patients. We modeled the physical decay of [123]I (half-life, 13.27 hours) and a constant biologic distribution of the isotope over the four-hour duration. The scanning time was the same in the first round of DaT SPECT scan. The SPECT scanners and acquisition parameters were identical to those in the first scan.

At the conclusion of the first round of scan for virtual patients generated from group A and group B, and the second round of scan for virtual patients generated from group B, we had realistically simulated a total of $N$ = 220 DaT SPECT studies (197 from the first scan and 23 from the second scan), with both SPECT scanners.

Table 1. The Acquisition Parameters of GE-scanner and Siemens-scanner

| SPECT scanner | GE-scanner | Siemens-scanner |
|---|---|---|

| Collimator type | Low energy high resolution | |
|---|---|---|
| Collimator grid | Parallel hole | |
| Energy resolution (@140 keV) | 9.8% | 9.9% |
| Crystal size | 40 cm×54 cm | 44.5 cm×59.1 cm |
| System resolution (FWHM@100 mm) | 7.4 mm | 8.0 mm |
| System sensitivity (@100 mm) | 72 cps/MBq | 91 cps/MBq |
| Number of views | 120 | |
| Orbit | 360 degrees | |
| Acquisition mode | Step and shoot | |
| Scatter-energy window | 90-143 KeV | |
| Photopeak-energy window | 143-175 KeV | |

*Regional DaT Uptake Quantification: DaT-CTLESS Method and Comparison Studies*

We divided the $N$ = 197 virtual patients into training ($N$ = 150) and test ($N$ = 47) datasets. The size of test dataset was determined by a power analysis that will be described later. The DaT-CTLESS method was trained and tested with SPECT projections obtained from GE-scanner. We first reconstructed photopeak-window projections using an ordered subset expectation maximization (OSEM)-based approach yielding non-AC activity maps. We used ten iterations and ten subsets in the OSEM-based reconstruction approach, as suggested previously[54]. The scatter-window projections were also reconstructed using the same OSEM-based approach, providing initial estimates of attenuation maps. These reconstructed images had a size of 128 × 128 × 128 with a voxel size of 0.35 cm, serving as the input for DaT-CTLESS method. The DaT-CTLESS method was trained on segmented CT-derived attenuation maps. The performance of DaT-CTLESS was evaluated on the clinical task of estimating DaT uptake within caudate, putamen, and GP regions.

In addition, we estimated the regional DaT uptake using CTAC and UAC. For CTAC, we reconstructed the photopeak-window projections using an OSEM-based approach with ten iterations and ten subsets and CT-derived attenuation maps for AC. For the UAC method, following the SNM Practice Guideline, we delineated an elliptical region over non-AC activity maps and assigned an attenuation coefficient of 0.11 cm$^{-1}$ to these regions, yielding uniform attenuation maps[43]. The UAC method used the OSEM-based approach with the same iteration and subset numbers as used in DaT-CTLESS and the uniform attenuation maps for AC.

We also compared the proposed method with two DL-based methods, one an indirect AC method and another a direct AC method. The indirect method, referred to as estDLAC, takes reconstructed images in scatter and photopeak windows as input and directly estimates attenuation coefficients within each voxel of the attenuation map. The estDLAC method generates attenuation maps by directly estimating the attenuation coefficient within each voxel from photopeak and scatter-window reconstructions. The estDLAC method was optimized on the mean-square-error loss between the estimated and CT-derived attenuation maps. The direct AC method, referred to as drctDLAC method, takes non-AC activity maps reconstructed in the photopeak-energy window as input and estimates activity maps with AC. The method directly estimates activity maps with AC from photopeak-energy window reconstructions. The drctDLAC method was optimized on the mean-square-error loss between the estimated and CTAC-derived activity maps. Both estDLAC and drctDLAC were based on U-net with structure similar to that for DaT-CTLESS, but with a rectified linear unit applied in the final layer. Also, the input of drctDLAC contained only one channel that takes the photopeak-energy window reconstructed images. To evaluate the performance of DaT-CTLESS, estDLAC, and drctDLAC with different sizes of the training dataset, we trained those methods with the number of training samples ranging from 50 to 150.

Except for the procedure to obtain the attenuation maps, all other reconstruction procedures were the same as those of the DaT-CTLESS method in the implementations of CTAC, UAC, and estDLAC methods.

As mentioned in the previous subsection, each patient was scanned by both GE-scanner and Siemens-scanner. To assess the generalizability across different SPECT scanners, the DaT-CTLESS method was also evaluated with test data from Siemens-scanner on the clinical task of estimating DaT uptake within caudate, putamen, and GP regions. Moreover, the DaT-CTLESS method was separately trained on samples scanned by Siemens-scanner and was evaluated with test data from both GE and Siemens scanners.

*Statistical Considerations*

For the primary objective, we calculated ICC between the regional DaT uptake estimated by the DaT-CTLESS method and those estimated by the CTAC method, as well as between those estimated by the UAC and CTAC methods. The ICC was calculated based on a two-way fixed single-score model for absolute agreement and interpreted according to the guideline proposed by Koo and Li[55]. To determine the sample size in the test dataset, we conducted a power analysis following the strategy proposed by Walter *et al.*[56]. We expect an ICC of at least 0.75 between DaT-CTLESS and CTAC. We found that $N$ = 33 samples were needed to detect an ICC under the proposed method of 0.90 against a null ICC of 0.75, using a 2-sided F-test at a significance level of 0.05 and for 80% power. Accordingly, we assigned $N$ = 47 virtual patients to the test dataset, comprising 24 generated from group A and 23 generated from group B. We calculated the ICC accompanied with a 95% confidence interval (CI) for DaT-CTLESS with CTAC as well as for UAC with CTAC, and for comparison, calculated the ICC difference with a 95% CI using a bootstrapping strategy.

We also calculated normalized RMSE (NRMSE) between the estimated regional DaT uptake and the ground truth. The NRMSE was normalized with respect to the ground truth DaT uptake. Based on these data, different AC methods were compared based on NRMSE difference between methods while the associated 95% CIs were constructed using a bootstrapping strategy. To account for multiple hypothesis testing, Bonferroni correction was applied.

In the test dataset, we have $N$ = 47 virtual patients, with $N$ = 24 having reduced SBR values and $N$ = 23 having normal SBR values. In a previous study, a significant difference was reported between the SBR value of putamen in PD patients and that in normal controls[48]. Thus, considering the use of SBR value to differentiate patients with PD from normal controls, we assessed the efficacy of the three AC methods (CTAC, DaT-CTLESS, and UAC) on the task of yielding SBR values that could perform this differentiation. For this purpose, we calculated the SBR values within the left and right putamen from the SPECT images reconstructed by three considered AC methods in our entire test dataset. Using the average of the SBR within the left and right putamen as a test statistic, we performed a receiver operating characteristic (ROC) analysis on the task of distinguishing patients with normal versus reduced SBR values and calculated the area under the ROC curve (AUC) for each of the three AC methods. We compared the paired AUC yielded by DaT-CTLESS and UAC on the same data by the Delong's test[57].

As previously mentioned, $N$ = 23 virtual patients generated from group B in the test dataset underwent retest scans. Based on these data, we evaluated the test-retest repeatability of DaT-CTLESS by calculating the within-subject coefficient of variation (WSCV). We grouped six VOIs into three VOI-groups, including caudate-group, putamen-group, and GP-group. WSCV values were calculated separately for these three VOI groups. In addition, we calculated normalized RMSE and structural similarity index measurement (SSIM) between images yielded by the DaT-CTLESS method and those from CTAC. Using a bootstrapping strategy, we assessed the statistical significance of the differences in these FoMs (RMSE and SSIM) between DaT-CTLESS and UAC. Further, to assess the sensitivity of DaT-CTLESS to heterogeneity within VOIs, we evaluated the performance of DaT-CTLESS with virtual patients where different levels of heterogeneity were modeled in patients with reduced SBR values. Finally, we compared the ICCs between CTAC and DaT-CTLESS, estDLAC and drctDLAC across different sizes of training dataset.

All statistical tests in ISIT-DaT were two sided and statistical significance was claimed with a $p$-value < 0.05.

## RESULTS

### Trial Accrual

The characteristics of the patients in ISIT-DaT cohort are shown in Table 2. A total of 197 virtual patients were included, where 174 were from group A and 23 were from group B. Virtual patients from group B underwent a secondary SPECT scan at a later time point. At each time point, each virtual patient was scanned using two different SPECT scanners. Consequently, a total of 220 DaT SPECT studies were generated.

Table 2. The Characteristics of Virtual Patients in ISIT-DaT Cohort

| | Characteristics | Summary[a] |
|---|---|---|
| **Group A** | Number of virtual patients | 174 |
| | Number of virtual patients with reduced SBR values | 87 (50%) |
| | Number of DaT SPECT studies | 174 |
| | Age | 70 (64.25, 75) |
| | Sex | 103 females (59.2%) |
| | Number of virtual patients in the training dataset | 150 (86%) |
| | Number of virtual patients in the test dataset | 24 (14%) |
| **Group B** | Number of virtual patients | 23 |
| | Number of virtual patients with reduced SBR values | 12 (52%) |
| | Number of DaT SPECT studies | 46 |
| | Age | 71 (68.5, 76.5) |
| | Sex | 11 females (47.8%) |
| | Number of virtual patients in the training dataset | 0 (0%) |
| | Number of virtual patients in the test dataset | 23 (100%) |

[a]Count (%) for categorial characteristics and median (interquartile range) for age.

**Regional DaT Uptake Quantification**

Table 3 shows the ICCs between the regional DaT uptake estimated by DaT-CTLESS vs. CTAC, as well as between those estimated by UAC vs. CTAC. The ICC between DaT-CTLESS vs. CTAC was observed to be significantly higher than ICC with UAC vs. CTAC ($p$<0.05). The ICCs between DaT-CTLESS and CTAC were around 0.96, indicating an excellent positive agreement[55].

Table 3. The ICCs between regional DaT uptake estimated by DaT-CTLESS and UAC with those estimated by CTAC

| Test data | AC method | ICC | 95% CI | |
|---|---|---|---|---|
| | | | Lower limit | Upper limit |
| GE-scanner | DaT-CTLESS | 0.96 | 0.94 | 0.97 |
| | UAC | 0.44 | 0.37 | 0.50 |

Fig. 4 shows normalized RMSE between the ground truth DaT uptake within the caudate, putamen, and GP regions and those obtained by CTAC, DaT-CTLESS, and UAC. For all VOIs, DaT-CTLESS performed similarly to the CTAC method and significantly outperformed the UAC method (p<0.05).

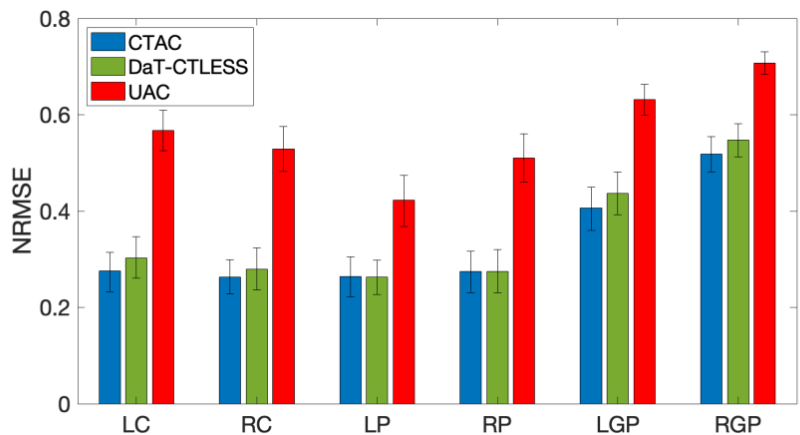

Figure 4. Normalized RMSE (NRMSE) of between the estimated DaT uptake and the ground truth within six VOIs. LC/RC, LP/RP, and LGP/RGP denote the left/right caudate, left/right putamen, and left/right globus pallidus, respectively. Error bars show the 95% CIs.

**Differentiating Patients with Normal and Reduced SBR Values**

Fig. 5 shows AUC values yielded by CTAC, DaT-CTLESS, and UAC. We observed that DaT-CTLESS yielded AUC values close to those obtained by using CTAC. Further, the AUC values obtained by DaT-CTLESS were significantly higher than those obtained using the UAC approach ($p<0.05$).

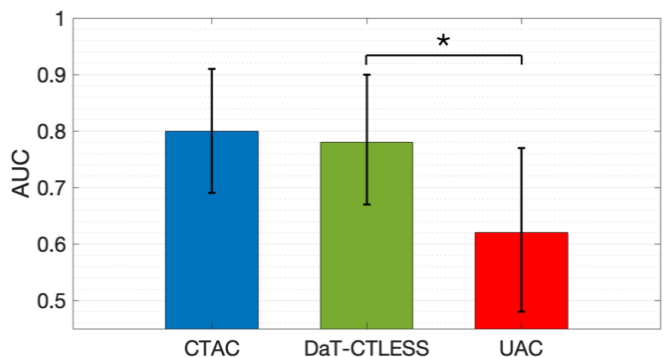

Figure 5. AUC yielded by CTAC, DaT-CTLESS, and UAC under two scanners. Stars indicate $p$-values less than 0.05.

**Test-retest Repeatability**

Fig. 6 shows the WSCV for three VOI-groups obtained by CTAC, DaT-CTLESS, and UAC in the test-retest study. Across all VOI-groups, DaT-CTLESS yielded WSCV values that were similar to those of CTAC for both SPECT scanners.

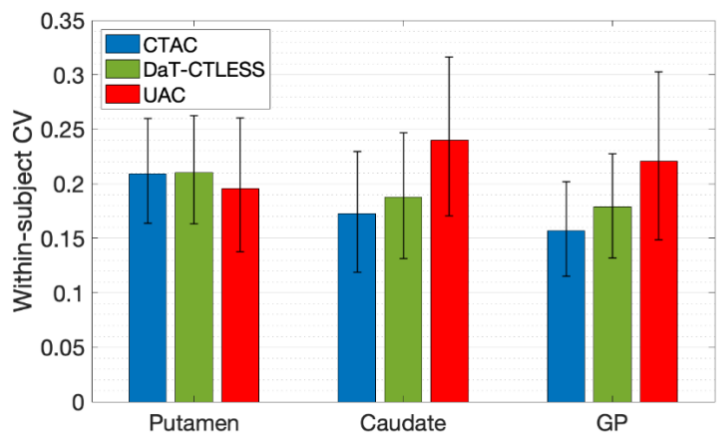

Figure 6. Within-subject CV calculated for each VOI-group in the test–retest experiment.

**Fidelity-based Similarity**

Fig. 7 presents the normalized RMSE and SSIM values comparing the activity and attenuation maps obtained by the DaT-CTLESS method with those obtained using the CTAC method. We also evaluated the performance of UAC on these FoMs for comparison. We observed that the DaT-CTLESS method significantly outperformed the UAC method on these fidelity-based FoMs for both SPECT scanners ($p<0.05$).

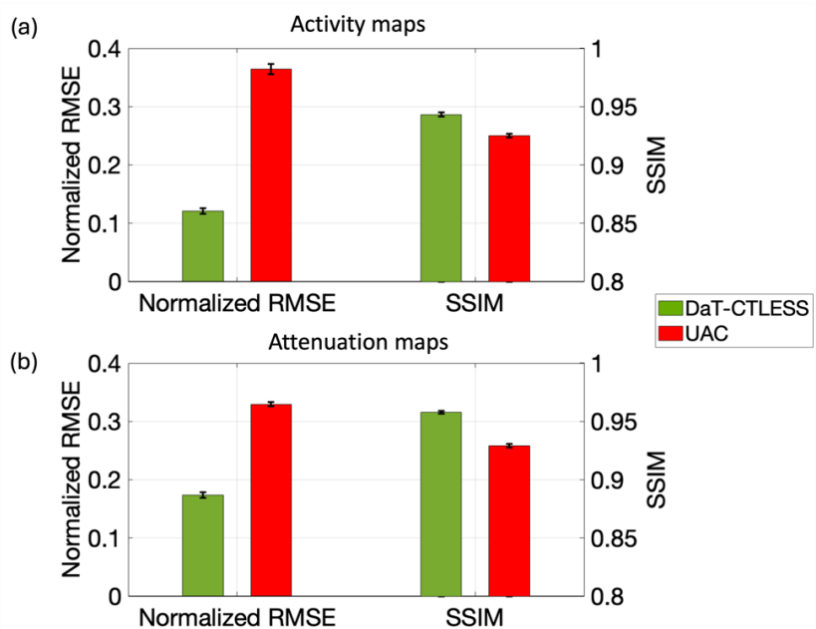

Figure 7. Normalized RMSE and SSIM between (a) attenuation maps as well as (b) activity maps obtained by the CTAC method and those obtained by DaT-CTLESS and UAC.

Fig. 8 shows representative examples of reconstructed activity maps and corresponding attenuation maps obtained using CTAC, DaT-CTLESS, and UAC. In Fig. 8a, the attenuation maps obtained by the DaT-CTLESS method visually looked similar to those derived from CT images. Note that the example in the second column of Fig. 8a contains the head gear, which was correctly estimated by DaT-CTLESS. Fig. 8b shows the reconstructed activity maps from a virtual patient with normal putamen SBR and a virtual patient with abnormal putamen SBR. The activity maps obtained by the DaT-CTLESS method looked similar to those obtained by CTAC method in both samples.

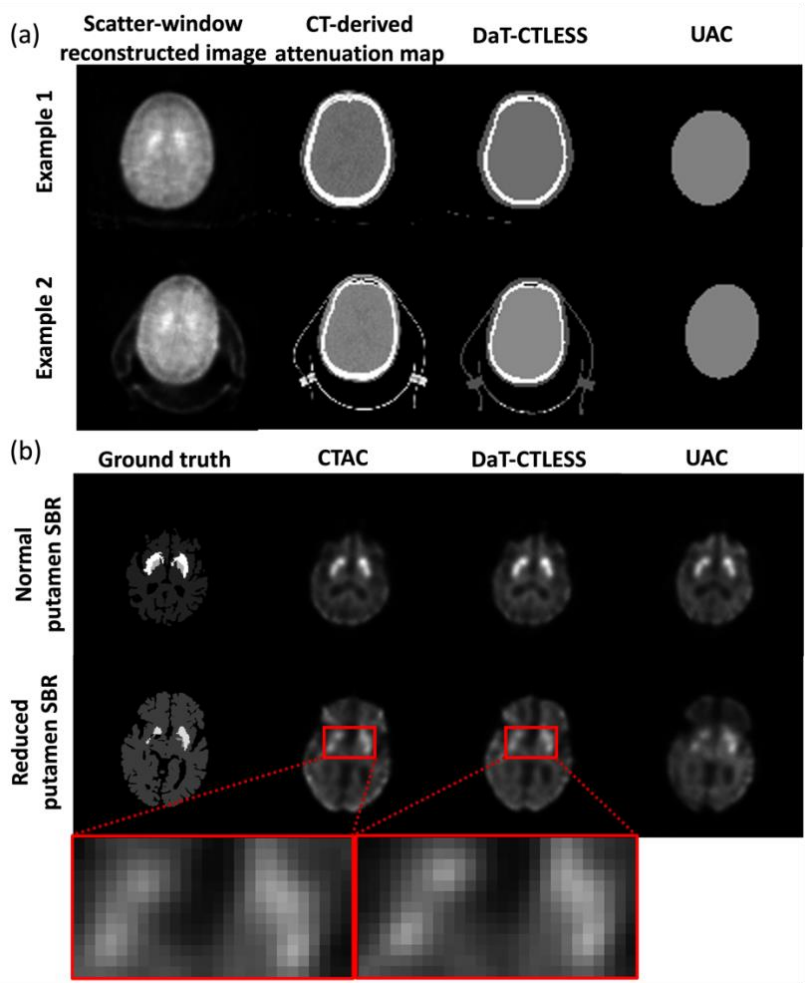

Figure 8. Representative examples of attenuation maps and reconstructed images. (a) Scatter-window reconstructed images, along with attenuation maps obtained by DaT-CTLESS and UAC, as well as derived from CT images. (b) Activity maps reconstructed by CTAC, DaT-CTLESS, and UAC. In (b), the upper example represents a virtual patient with normal putamen SBR, and the lower example represents a virtual patient with reduced putamen SBR. Zoomed insets were given in the red-boxed regions.

**Sensitivity to Intra-regional Heterogeneity**

Fig. 9 shows the ICCs yielded by DaT-CTLESS and UAC across different levels of heterogeneity within VOIs, as generated by varying the weight of the heterogeneity, $\mathcal{W}$, in Eq. 7. We observed that for all values of $\mathcal{W}$, DaT-CTLESS yielded significantly higher ICC values than UAC ($p$<0.05).

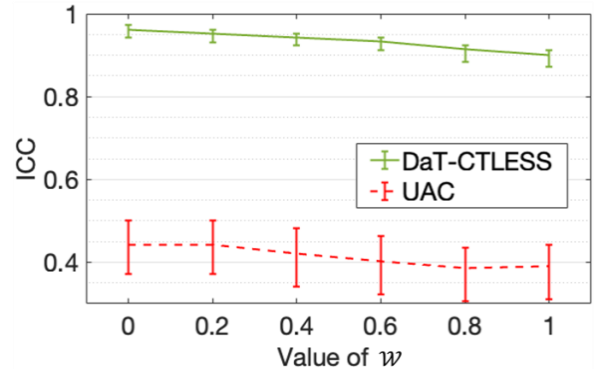

Figure 9. The ICCs between the DaT uptake obtained by CTAC and those obtained by DaT-CTLESS and UAC with different values of $\mathcal{W}$.

**Evaluation with Different Sizes of Training Dataset**

Fig. 10 shows the ICCs between the regional DaT uptake obtained by CTAC and those obtained by DaT-CTLESS, estDLAC and drctDLAC for different sizes of the training dataset. We observed that the DaT-CTLESS method generally yielded higher ICCs than estDLAC. Additionally, while the performance of estDLAC degraded when the number of training samples reduced below 100, the DaT-CTLESS method still maintained a good alignment with the CTAC method and significantly outperformed the estDLAC method. In fact, even with only 50 patients, the DaT-CTLESS method was still close to the CTAC method with an ICC value above 0.95 on the regional DaT uptake quantification task. Further, we observed that the DaT-CTLESS method yielded significantly higher ICCs than the drctDLAC method across different sizes of training dataset.

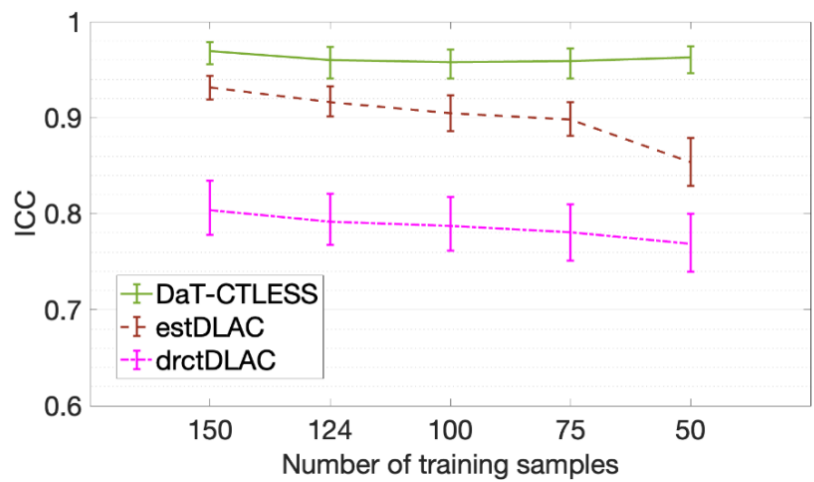

Figure 10. The ICCs between the DaT uptake estimated by CTAC and those estimated by DaT-CTLESS, estDLAC, and drctDLAC with different sizes of training dataset.

**Assessing generalizability of the DaT-CTLESS method**

To assess the generalizability across different SPECT scanners, the DaT-CTLESS method was separately trained on samples scanned by two considered SPECT scanners. We refer to DaT-CTLESS trained on data from the GE scanner as DaT-CTLESS-GE, and DaT-CTLESS trained on data from the Siemens scanner as DaT-CTLESS-SMS. We evaluated the performance of DaT-CTLESS-GE and DaT-CTLESS-SMS with test data from GE-scanner as well as from Siemens-scanner.

Table 4 presents the ICCs between regional DaT uptake estimates obtained using CTAC and those derived from UAC, DaT-CTLESS-GE, and DaT-CTLESS-SMS. For both the considered scanners, the ICCs between DaT-CTLESS vs. CTAC were observed to be significantly higher than ICC with UAC vs. CTAC ($p<0.05$).

Table 4. The ICCs between regional DaT uptake estimated by DaT-CTLESS and UAC with those estimated by CTAC with data from different scanners.

| Test data | AC method | ICC | 95% CI | |
|---|---|---|---|---|
| | | | Lower limit | Upper limit |
| GE-scanner | DaT-CTLESS-GE | 0.96 | 0.94 | 0.97 |
| | DaT-CTLESS-SMS | 0.96 | 0.94 | 0.97 |
| | UAC | 0.44 | 0.37 | 0.50 |

| Siemens-scanner | DaT-CTLESS-GE | 0.96 | 0.95 | 0.97 |
| | DaT-CTLESS-SMS | 0.97 | 0.95 | 0.98 |
| | UAC | 0.43 | 0.37 | 0.50 |

Fig. 11 shows normalized RMSE between the ground truth DaT uptake within the caudate, putamen, and GP regions and those obtained by considered AC methods. For all the VOIs and the two considered scanners, DaT-CTLESS performed similarly to the CTAC method and significantly outperformed the UAC method ($p<0.05$).

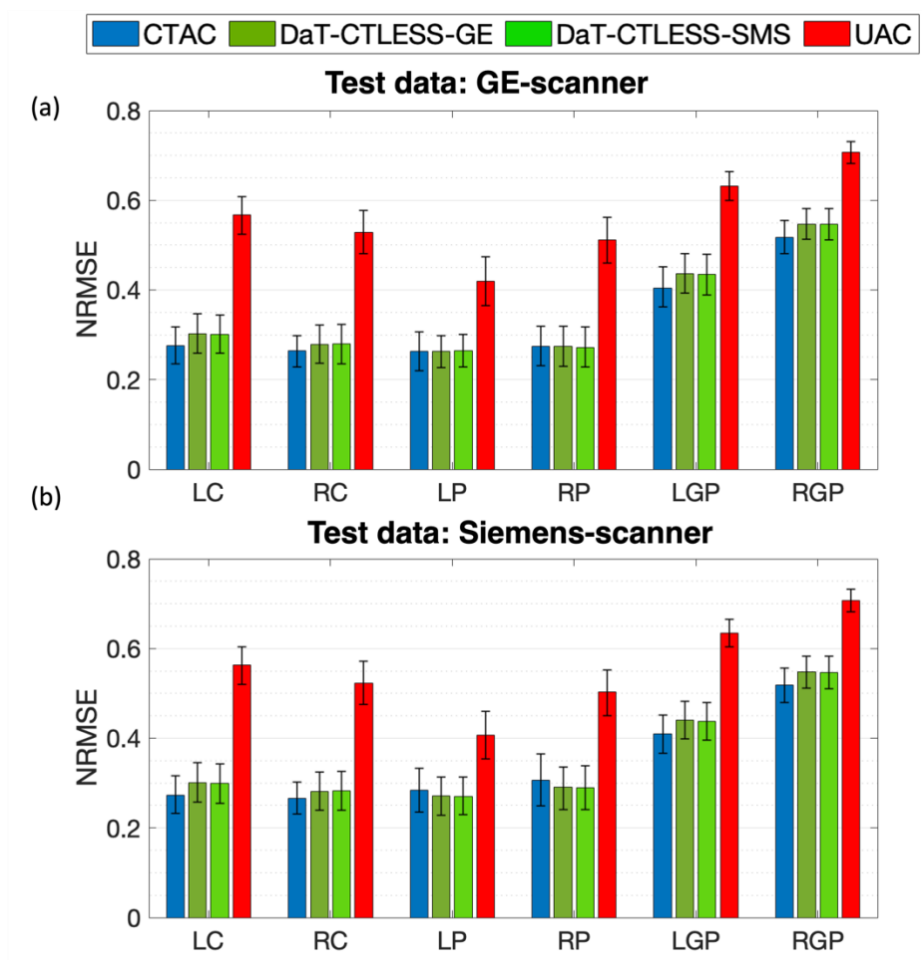

Figure 11. Normalized RMSE (NRMSE) between the estimated DaT uptake and the ground truth within six VOIs. LC/RC, LP/RP, and LGP/RGP denote the left/right caudate, left/right putamen, and left/right globus pallidus, respectively. Error bars show the 95% CIs. Test data were obtained from (a) GE-scanner and (b) Siemens-scanner.

**DISCUSSION**

Reliable quantification of DaT uptake in DaT SPECT images requires performing AC, which typically requires an attenuation map. These maps are usually obtained either by conducting a separate transmission scan, or by delineating an elliptical region over the activity reconstruction and assigning a fixed attenuation coefficient value. The former approach typically involves a separate CT scan and presents several disadvantages as discussed in the Introduction section, and the latter has shown limited performance[58,59]. To address these challenges, we proposed a scatter-window projection and deep learning-based transmission-less attenuation compensation method for DaT SPECT, namely DaT-CTLESS. We conducted an *in-silico* imaging trial, namely ISIT-DaT, to evaluate the DaT-CTLESS method on the clinical task of regional DaT uptake quantification.

In ISIT-DaT, we observed an excellent positive agreement between the DaT-CTLESS method and the CTAC method on the task of regional activity quantification with an ICC of 0.96, while the ICC between the UAC and CTAC methods was 0.43 (Table 3). In addition, as shown in Fig. 4, the DaT-CTLESS method yielded a similar performance to the CTAC method in terms of normalized RMSE relative to ground truth, and significantly outperformed the UAC method across all VOIs on the regional DaT uptake quantification task. The ability of performing AC without CT scan while maintaining reliable regional DaT uptake quantification performance demonstrated by DaT-CTLESS has important clinical implications. First, the DaT-CTLESS method could offer a direct advantage for DaT SPECT studies conducted on SPECT-only scanners. As mentioned previously, a significant portion of SPECT systems do not include CT components. The DaT-CTLESS method could provide AC without the need of CT in those systems, thus, making DaT SPECT studies with AC at potential lower cost and available to larger numbers of patients. Second, the DaT-CTLESS method inherently avoids the potential of quantification error caused by misalignment between SPECT and CT scans. Resting tremor and medication-related dyskinesia are commonly associated in patients with PD[43], leading to high occurrences of misalignment between SPECT and CT scans in DaT SPECT studies. Since any patient motion that occurs during the scan equally affects the photopeak and scatter window data, the DaT-CTLESS method is expected to be insensitive to such misalignment issues. Further, the DaT-CTLESS method offers advantages in terms of reduced radiation dose. While CT scans for AC are typically administered at low dose, there is a concern regarding the potential cancer risk associated with even low levels of radiation exposure[60]. The *as low as reasonably achievable* (ALARA) principles advocate making every reasonable effort to keep ionizing radiation exposures well below practical dose limits, taking various factors into account, including technological capabilities[61]. Thus, the DaT-CTLESS method is favored as an imaging technique that yields quantification performance with excellent positive agreement to the CTAC method while eliminating the radiation exposure from CT scans.

Our results show reliable performance of the DaT-CTLESS method in the task of regional uptake quantification. However, eventually, regional uptake is used to distinguishing PD patients from healthy controls. Thus, a more directly and clinically relevant evaluation is to examine the discriminative ability of DaT-CTLESS in separating these patient populations in comparison to CTAC and UAC methods. Such an evaluation would however require knowledge of the PD ground truth, which is challenging as the only diagnostic confirmation for PD is histopathological examination. Currently, no clinical examination, imaging study, or other diagnostic markers can achieve complete accuracy. In this context, we recognized that PD is the most common etiology of striatal dopamine deficiency. Notably, Son *et al.* observed a significantly lower SBR value of putamen in PD patients compared to that in normal controls[48]. Based on this observation on the difference in SBR values, we evaluated DaT-CTLESS on the task of differentiating patients between normal and reduced SBR values. We observed that DaT-CTLESS yielded similar performance as CTAC on this task. Further, DaT-CTLESS significantly outperformed the UAC method, indicating that when the CT scans are unavailable, the DaT-CTLESS method may provide a more accurate way to distinguish between patients with normal vs. low SBR values. The promising performance of DaT-CTLESS observed in ISIT-DaT motivates further clinical evaluation assessing its role in diagnosing PD.

In the test–retest experiment (Fig. 6), DaT-CTLESS yielded WSCV values comparable to those of CTAC, indicating that DaT-CTLESS offers a repeatability similar to CTAC. This result suggests that DaT-CTLESS is only as sensitive to imaging system noise as CTAC and is clinically significant[62]. Reliable tracking of DaT uptake in patients with PD over time is essential for monitoring disease progression and evaluating therapeutic response. If an AC method yields inconsistent uptake values across repeated scans, it could lead to misleading clinical interpretations. By demonstrating a similar test–retest repeatability to CTAC, DaT-CTLESS supports its potential for clinical translation.

Fig. 7 shows that the DaT-CTLESS method significantly outperformed the UAC method on FoMs of RMSE and SSIM, when comparing attenuation and activity maps with those obtained by the CTAC method as a reference. In addition, as shown by an example in Fig. 8, DaT-CTLESS generated activity maps and attenuation maps look visually similar to those obtained by the CTAC method. Importantly, DaT-CTLESS reliably estimated the attenuation distribution in areas without activity, such as regions affected by head gear (Fig. 8a), a capability that the UAC method inherently lacks. A previous study found that head gear attenuation led to underestimation of DaT uptake in the parietal and occipital cortices, as well as the cerebellum, in activity maps reconstructed by the UAC method[6]. Thus, addressing the attenuation effect caused by head gear is of much clinical importance for accurate DaT SPECT.

Dopamine distribution in the striatum of patients with Parkinson's disease may exhibit intra-regional heterogeneity[63]. To assess the sensitivity of DaT-CTLESS to such intra-regional heterogeneity, we simulated varying levels of heterogeneity in activity distributions within VOIs for patients with reduced SBR values. As shown in Fig. 9, despite being trained on data with homogeneously distributed activity within the VOIs, DaT-CTLESS constantly yielded significantly higher ICC values than UAC across all levels of the intra-regional heterogeneity. These observations suggest that DaT-CTLESS is relatively insensitive to intra-regional heterogeneity and support its potential for clinical evaluation.

One well-known challenge in the application of DL-based medical imaging methods is the limited availability of training data. In this context, we observed that the DaT-CTLESS method maintained a good performance on the quantification task with different sizes of training data, even with as few as $N = 50$ samples. Moreover, as shown Fig. 10, the DaT-CTLESS method yielded higher ICCs than two other DL-based methods, one indirect (estDLAC) and another direct (drctDLAC) with different sizes of the training data. Notably, while estDLAC is also an indirect DL-based AC method, DaT-CTLESS demonstrated superior performance. To explain this finding, we recognize that the correlation between SPECT emission data and attenuation maps for each voxel is complex and varies across regions. Thus, a data-driven approach that directly estimates attenuation coefficients within each voxel can require a large amount of training data. The DaT-CTLESS method addresses this issue by leveraging the fact that attenuation coefficients are relatively uniform and approximately known within specific regions of the head. Using this prior information, the DaT-CTLESS method posits the determination of attenuation map as a segmentation problem, instead of estimating attenuation coefficient of each voxel. This approach reduces the dimensionality of the task, helping reduce the requirement for large training data, thereby outperforming the estDLAC method with the same size of training dataset.

Clinical adoption of DL-based medical-imaging methods requires their ability to generalize across different clinical scanners[39]. This is important, as lack of generalizability would necessitate retraining the method for each scanner, thereby making its application impractical. In silico imaging trials provide an inexpensive and easier approach to assess generalizability. In ISIT-DaT, we evaluated the DaT-CTLESS method on SPECT scanners from two different vendors. More specifically, the DaT-CTLESS method was independently trained and tested on data from a GE Discovery 670 scanner and a Siemens Symbia Evo Excel scanner. As shown in Table 4 and Fig. 11, across those two considered scanners, the DaT-CTLESS method yielded an excellent positive agreement to the CTAC method on the task of regional activity quantification with an ICC of 0.96 and significantly outperformed UAC method. These findings demonstrate strong promise of the generalizability of DaT-CTLESS and motivate studies to assess the performance of DaT-CTLESS in a multicenter trial.

The ISIT-DaT trial was meticulously designed to ensure a high degree of clinical realism throughout the trial pipeline, including a careful assembly of the patient cohort and accurate simulation of the SPECT systems. The anatomical characteristics of the virtual patients were derived from real MR and CT images, while the physiological characteristics, including the ratio of striatal to occipital [123]I-ioflupane binding based on clinical data of PD patients and non-PD individuals with variable values in measured VOIs[48]. The SPECT systems were modeled using well-validated Monte Carlo simulations, with clinically relevant parameters and the imaging protocol aligned with established clinical guidelines[43]. The inclusion of SPECT systems from different vendors allowed for a comprehensive evaluation of DaT-CTLESS, replicating the variability encountered in real-world practice. Notably, we evaluated the performance of DaT-CTLESS on the clinical task of regional DaT uptake quantification and distinguishing patients with normal versus with reduced SBR values, making evaluation encouraging for clinical relevance and impact.

Although the ISIT-DaT trial closely modeled clinical conditions and provided a mechanism to conduct evaluations through an approach that was inexpensive and logistically practical, there were also some tradeoffs given the absence of clinical data. For example, in modeling the patient physiology, we considered a static distribution of the tracer across acquisitions and considered only physical decay of the tracer in virtual patients due to the limited dataset of patient-specific [123]I pharmacokinetics. A promising direction for future research is to incorporate models of pharmacokinetics into virtual patient simulations. Similarly, in the context of a multi-center evaluation, we focused on modeling variations in imaging systems. However, ISIT-DaT did not simulate other sources of variabilities, including variability in patient populations from different geographical locations. Thus, real-world validation of findings in ISIT-DaT is an important future direction prior to real-world clinical application of DaT-CTLESS.

The promising outcomes observed in ISIT-DaT motivate clinical trials for evaluation of DaT-CTLESS. We anticipate that clinical trials will demonstrate reliable performance for DaT-CTLESS, given the excellent positive agreement to CTAC in regional DaT uptake quantification, good discrimination between patients with normal versus with reduced SBR values, high fidelity-based similarity in generated SPECT images, strong generalizability across two different SPECT scanners, similar test-retest repeatability to CTAC, and relatively low requirement for the number of training samples. This would provide more compelling evidence for the clinical implementation of DaT-CTLESS in transmission-less AC for DaT SPECT.

## CONCLUSION

We proposed a scatter-window projection and learning-based transmission-less attenuation compensation (AC) method for DaT SPECT (DaT-CTLESS) and evaluated the performance of the method in an *in-silico* imaging trial (ISIT-DaT). Results from ISIT-DaT demonstrated that the proposed method yielded similar performance as a CT-based AC method (CTAC) and significantly outperformed an AC method with uniform attenuation map (UAC) on the task of regional uptake quantification. Additionally, DaT-CTLESS yielded significantly higher ICC values than UAC across different levels of intra-regional heterogeneity. Further, the DaT-CTLESS method was observed to significantly outperform the UAC method on the task of distinguishing patients with normal versus with reduced putamen specific binding ratios. The proposed method generalized well across SPECT scanners from two vendors and yielded similar test-retest repeatability to CTAC. Finally, as we reduced the size of the training data set, the DaT-CTLESS method yielded relatively stable performance and outperformed an indirect deep learning-based AC approach and another direct deep learning-based AC approach.  As per the RELAINCE guidelines, ISIT-DaT generated the following claim:

*A scatter-window projection and learning-based transmission-less attenuation compensation method for DaT SPECT (DaT-CTLESS) yielded a significantly higher correlation with a CT-based AC method than the correlation between a uniform attenuation map AC method and the CT-based AC*

*method on the task of quantifying DaT uptake in the caudate, putamen, and globus pallidus as evaluated in a virtual imaging trial with single-center multi-scanner data. The intra-class correlation coefficient of the estimated regional uptake obtained by the DaT-CTLESS method and those obtained using the CT-based AC method was 0.96 with 95% confidence interval of [0.94, 0.97].*

These results demonstrate the capability of DaT-CTLESS method for transmission-less AC in DaT SPECT and provide strong motivation for clinical evaluation of DaT-CTLESS, including in multi-center settings.

## ACKNOWLEDGMENTS

This work was supported in part by National Institute of Health with support from grants R01-EB031051, R01-EB031962, and R01-NS124789.

## CONFLICT OF INTEREST STATEMENT

The authors have no relevant conflicts of interest to disclose.